\documentclass{article}
\usepackage{amsmath}
\usepackage{graphicx}
\begin{document}
\centerline
{\large \bf The Tsallis Distribution at the LHC.}
\medskip
\vskip 1cm

\centerline{\bf J. Cleymans}
\vskip 1cm
\begin{center}
UCT-CERN Research Centre and Department of Physics, \\
University of Cape Town, Rondebosch, South Africa  
\end{center}
\vskip 1.5cm

\begin{abstract}
The Tsallis distribution has been used widely in high energy physics to describe the transverse 
momnetum distributions of particles. In this note we show that the use of a thermodynamically consistent
form of this distribution leads to a description of identified particles with the same values of the 
temperature $T$ and the parameter $q$.
\end{abstract}

There exists a rich and wide variety of distributions covering a large range of applications~\cite{newman,clauset}.
Those having a power law behaviour have attracted considerable attention in physics in 
recent years but there is a a long history 
in other fieds such as biology and economics~\cite{mitzenmacher}.

In high energy physics power law distributions have been applied by a very large number
of scientists~\cite{STAR,PHENIX,ATLAS,ALICE,CMS} to the description of transverse momenta  
of secondary particles produced in $p-p$ collisions. Indeed the available 
range of transverse momenta  has expanded
considerably  with the advent of the Large Hadron Collider (LHC). 
Collider  energies up to 8 TeV are now available in $p-p$ collisions and 
transverse momenta of hundreds of GeV are now a common occurrence.
In this presentation the focus will be on various forms of distributions first proposed
by C. Tsallis about  twenty-five years ago~\cite{tsallis1}. 

In the analysis of the new data, a Tsallis-like distribution 
gives excellent fits to the transverse momentum 
distributions  as shown by the  
by  the ALICE~\cite{ALICE}, ATLAS~\cite{ATLAS} and CMS~\cite{CMS} collaborations at the LHC
and by the STAR~\cite{STAR} and PHENIX~\cite{PHENIX} collaborations at RHIC. 
In this paper we review the parameterization used by these groups and propose a slightly different one
which has a more consistent interpretation and has the bonus of being  thermodynamically consistent.

For high energy physics a consistent form  of Tsallis 
statistics~(see e.g.~\cite{worku2} and references therein) for 
the particle number, energy density and pressure is given by the
expressions given below 
\begin{eqnarray}
N &=& gV\int\frac{d^3p}{(2\pi)^3}
\left[ 1 + (q-1) \frac{E-\mu}{T}\right]^{-\frac{q}{q-1}} ,\label{number} \\
\epsilon &=& g\int\frac{d^3p}{(2\pi)^3}E
\left[ 1 + (q-1) \frac{E-\mu}{T}\right]^{-\frac{q}{q-1}} ,\label{epsilon}\\
P &=& g\int\frac{d^3p}{(2\pi)^3}\frac{p^2}{3E}
\left[ 1 + (q-1) \frac{E-\mu}{T}\right]^{-\frac{q}{q-1}}\label{pressure} .
\end{eqnarray}
where $T$ and $\mu$ are the temperature and the chemical potential,
$V$ is the volume and  $g$ is the degeneracy factor.
As is well-known the Tsallis distribution~\cite{tsallis1,tsallis2} introduces a new parameter $q$ which 
for transverse momentum spectra is  always close to 1,
typical values for the parameter $q$  obtained from fits to the transverse momentum distribution 
are in the range 1.1 to 1.2. In the remainder of this paper we will 
always assume $q>1$.\\
The expressions~\eqref{number},\eqref{epsilon} and \eqref{pressure} are  thermodynamically consistent, 
e.g. it can be easily shown~\cite{worku2} that  relations
of the type
\begin{equation}
 N = V\left.\frac{\partial P}{\partial \mu}\right|_{T},\label{consistency}
\end{equation}
are  satisfied~\cite{worku1,worku2}.
Note that without the extra power of $q$ in the equations~\eqref{number},\eqref{epsilon}, \eqref{pressure} the thermodynamic consistency 
would not be achieved.\\
It follows from~\eqref{number} that the momentum distribution is given by,
\begin{equation}
\frac{d^{3}N}{d^3p} = 
\frac{gV}{(2\pi)^3}
\left[1+(q-1)\frac{E -\mu}{T}\right]^{-q/(q-1)},
\label{tsallismu}
\end{equation}
or, expressed in terms of transverse momentum, $p_T$,  
transverse mass, $m_T$, and  rapidity  $y$  
\begin{equation}
\frac{d^{2}N}{dp_T~dy} = 
gV\frac{p_Tm_T\cosh y}{(2\pi)^2}
\left[1+(q-1)\frac{m_T\cosh y -\mu}{T}\right]^{-q/(q-1)},
\label{tsallismu1}
\end{equation}
At mid-rapidity $y=0$ and for zero chemical potential $\mu=0$ this reduces to
\begin{equation}
\left.\frac{d^{2}N}{dp_T~dy}\right|_{y=0} = 
gV\frac{p_Tm_T}{(2\pi)^2}
\left[1+(q-1)\frac{m_T}{T}\right]^{-q/(q-1)},
\label{tsallisLHC}
\end{equation}
This is the expression  used in~\cite{worku1,worku2} to fit the LHC
transverse momentum spectra.\\
It is well-known since 1988~\cite{tsallis1} that in the limit where the parameter $q$ goes to 1 this reduces 
to the standard Boltzmann distribution:
\begin{equation}
\lim_{q\rightarrow 1}\frac{d^{2}N}{dp_T~dy} = 
gV\frac{p_Tm_T\cosh y}{(2\pi)^2}
\exp\left(-\frac{m_T\cosh y -\mu}{T}\right).
\label{boltzmann}
\end{equation}
The parameterization given in Eq.~\eqref{tsallisLHC} is close to
the one used (but differernt) e.g. by the  ALICE~\cite{ALICE}, ATLAS~\cite{ATLAS}, CMS~\cite{CMS}, STAR~\cite{STAR} and 
PHENIX~\cite{PHENIX} collaborations where the following  form  is used : 
\begin{equation}
  \frac{d^2N}{dp_T\,dy} = p_T \frac{dN}
  {dy} \frac{(n-1)(n-2)}{nC(nC + m_{0} (n-2))}
 \left[ 1 + \frac{m_T - m_{0}}{nC} \right]^{-n}
\label{alice}
\end{equation}
where $n$, $C$ and $m_0$ are fit parameters. Indeed, after substituting 
\begin{equation}
n\rightarrow \frac{q}{q-1}
\label{n}
\end{equation}
and 
\begin{equation}
nC  \rightarrow \frac{T+m_0(q-1)}{q-1}  .
\label{nC}
\end{equation}
The Eq.~\eqref{alice} becomes
\begin{eqnarray}
  \frac{d^2N}{dp_T\,dy} =&& p_{T} \frac{{\rm d}N}
  {{\rm d}y} \frac{(n-1)(n-2)}{nC(nC + m_{0} (n-2))}\nonumber\\ 
&&\left[\frac{T}{T+m_0(q-1)}\right]^{-q/(q-1)}\nonumber\\
&&\left[ 1 + (q-1)\frac{m_T}{T} \right]^{-q/(q-1)}  .
\label{alice2}
\end{eqnarray}
Which, at mid-rapidity $y=0$ and zero chemical potential,
 has the same dependence on the 
transverse momentum as~\eqref{tsallisLHC} apart from an
additional  factor $m_T$ on the right-hand.
It has to be pointed out explicitly that the inclusion of the rest mass in the substitution Eq.~\eqref{nC}
is not in agreement with the Tsallis distribution as it breaks 
$m_T$ scaling which is present in the Tsallis form~\eqref{tsallismu1}
 but not in Eq.~\eqref{alice}. \\
The inclusions of the factor $m_T$ 
leads to a more consistent interpretation of the 
variables $q$ and $T$~\cite{worku1,worku2}. 
\\
The distribution~\eqref{tsallisLHC} has been used to fit the data
for identified particles, $\pi , K$ and $p$ for the ALICE~\cite{ALICE} 
collaboration  and $K^0_s$, $\Lambda$ and $\Xi$ for  the 
CMS~\cite{CMS} collaboration in $p-p$ collisions
at 900 GeV~\cite{worku1,worku2}.  The results are shown in Table 1 for the 
parameters $T$ and $q$.   The coresponding transverse
momentum distributions for the ALICE~\cite{ALICE}  are shown in Fig.~(1).  
For all identified particles the results are consistent with having
a system at  a Tsallis freeze-out temperature of about 
\begin{equation}
T \approx 70 \mathrm{MeV}
\end{equation}
and a value for the $q$ parameter of about  
\begin{equation}
q \approx 1.15
\end{equation}
These values are comparable to the ones obtained recently in~\cite{deppman,sena} where 
the original proposal of Hagedorn~\cite{hagedorn} was extended to a Hagedorn-Tsallis distribution.
\begin{figure}
\centering
\includegraphics[width=0.99\linewidth]{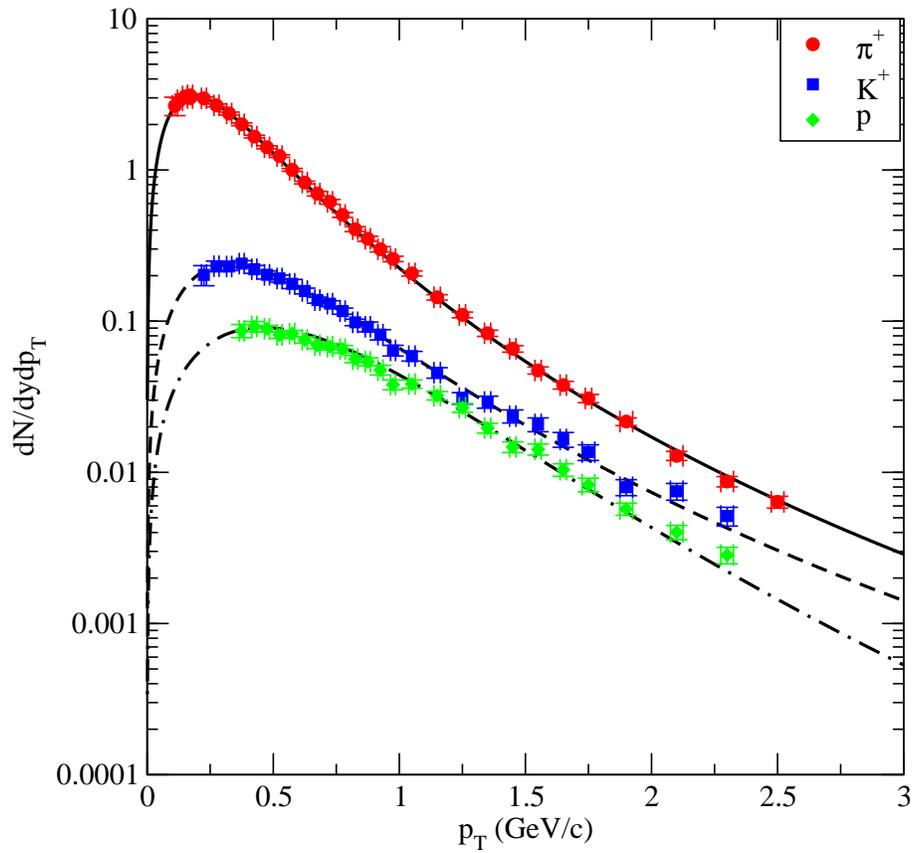}
\caption{Fit to the $\pi, K, p$ transverse momentum distributions
in $p-p$ collisions as measured by the ALICE collaboration~\cite{ALICE} using the Tsallis distribution 
function as given by \eqref{tsallisLHC}. }
\label{fig-1}       
\end{figure}


\begin{table}[t]
\begin{center}
\begin{tabular}{|l|c|c|}
\hline
\multicolumn{3}{|c|}{$p-p$ } \\
\multicolumn{3}{|c|}{ 900 GeV} \\
\hline
Particle & $q$            & $T$     \\
\hline 
$\pi^+$  & 1.154  $\pm$0.036 & 0.0682 $\pm $0.0026 \\ 
$\pi^-$  & 1.146  $\pm$0.036 & 0.0704 $\pm$ 0.0027 \\
$K^+$    & 1.158  $\pm$0.142 & 0.0690 $\pm $0.0223  \\
$K^-$    & 1.157  $\pm$0.139 & 0.0681 $\pm$ 0.0217  \\
$K^0_S$  & 1.134  $\pm$0.079 & 0.0923 $\pm $0.0139  \\
$p$      & 1.107  $\pm$0.147 & 0.0730 $\pm$ 0.0425   \\
$\bar{p}$& 1.106  $\pm$0.158 & 0.0764 $\pm $0.0464  \\
$\Lambda$& 1.114  $\pm$0.047 & 0.0698 $\pm$ 0.0148  \\
$\Xi^-$  & 1.110  $\pm$0.218 & 0.0440 $\pm$ 0.0752  \\
\hline  
\end{tabular}
\caption{Fitted values of the $T$ and $q$ parameters  
measured in $p-p$ collisions by the ALICE and CMS collaborations
 using the Tsallis form~\eqref{tsallisLHC} 
for the momentum distribution. 
}
\end{center}
\end{table}
%
%
\begin{figure}
\centering
\includegraphics[width=\textwidth,height=10cm]{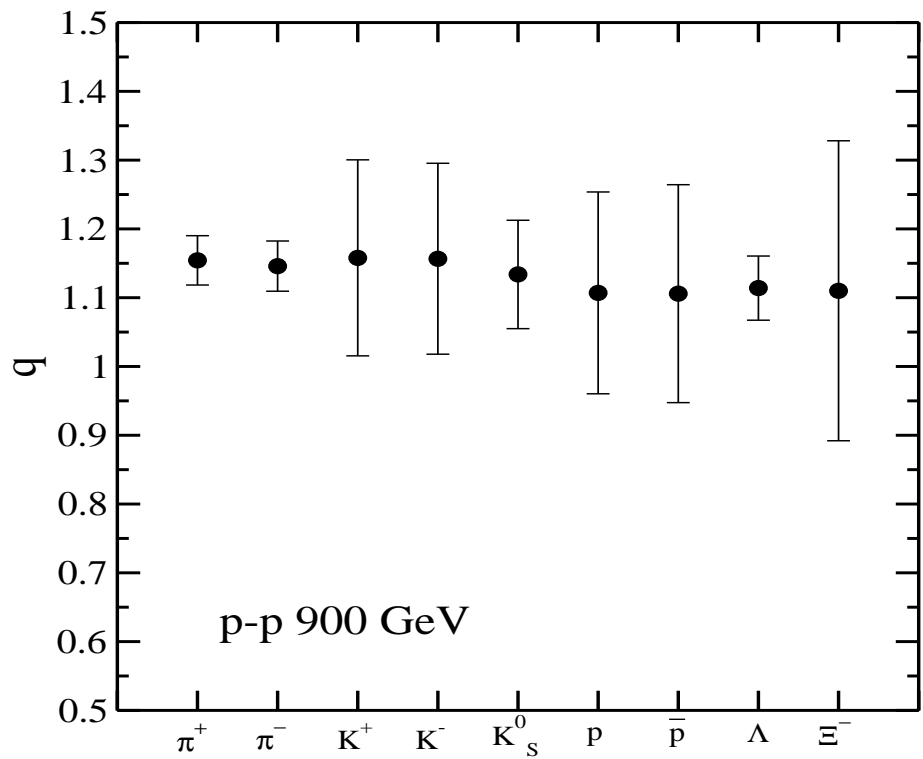}
\label{qfig}
\caption{Values of the Tsallis parameter $q$ for different  species of hadrons.}
\end{figure} 
%
The consistency of the values of $q$ is shown in Fig.~(2).\\
In conclusion we can say that the  use of the Tsallis parameterization
presented in~\eqref{number} leads to a good description 
of identified particles in $p-p$ collisions at 900 GeV with a consistent set of parameters.

\end{document}